\documentclass[a4paper,11pt]{article}
\usepackage{pos}

\title{Stereo observations of CTA~1 with SST-1M}
\author*[a]{ 
    B.~Lacave
}
\onbehalf{on behalf of the SST-1M Collaboration}

\affiliation[a]{Département de Physique Nucléaire et Corpusculaire, Université de Genève, Faculté de Sciences,\\
1211 Genève 4, Switzerland}

\emailAdd{bastien.lacave@unige.ch}

\abstract{CTA~1 is a composite supernova remnant featuring a shell structure and an inner Pulsar Wind Nebula. The shell is visible in the radio band, while Fermi has detected the radio-quiet pulsar PSR J0007+7303 at its core. Gamma-ray detectors such as LHAASO and VERITAS have detected TeV emission in the vicinity of the pulsar. However, the derived SEDs from LHAASO WCDA and VERITAS show significant discrepancies, which could be due to a complicated energy-dependent morphology not accounted for in the spectral analysis, and different angular resolution of the two experiments.
CTA~1 has been a target for dedicated observations by the SST-1M telescopes, a pair of small-sized Imaging Atmospheric Cherenkov Telescopes (IACTs) capable of operating in both mono and stereo modes. Located at the Ondřejov Observatory in Czech Republic, these telescopes are sensitive to the high energy range of the gamma-ray spectrum, spanning from 1 to 300 TeV. To investigate the very high-energy emission of CTA~1, the SST-1Ms have accumulated approximately 30 hours of selected observations, aiming to further constrain the characteristics of the source's high energy emission, and to shed some light into the discrepancy between different experiments.\cite{2011A&A...535A..64S}}

\ConferenceLogo{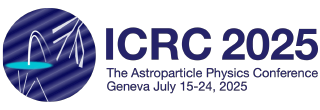}

\FullConference{39th International Cosmic Ray Conference (ICRC2025)\\
 15–24 July 2025\\
Geneva, Switzerland\\}
\begin{document}
\maketitle

\section{Introduction}

The composite supernova remnant (SNR) CTA~1 (G119.5+10.2) has recently sparked attention folowing the latest LHAASO results~\cite{2025SCPMA..6879503L} compared to VERITAS observations~\cite{2013ApJ...764...38A}. The remnant has a shell structure that is visible in radio wavelengths \cite{1997A&A...324.1152P}. At its center is a Pulsar Wind Nebula (PWN) powered by the radio-quiet pulsar PSR J0007+7303, which was discovered by the Fermi Large Area Telescope (LAT) \cite{2009Sci...325..840A, 2012ApJ...744..146A}. This inner nebula has also been clearly observed in X-rays \cite{2004ApJ...612..398H}.

At very high energies (VHE), CTA~1 is a known gamma-ray source. The VERITAS collaboration first detected it as an extended source named VER J0006+729 \cite{2013ApJ...764...38A}. More recently, the LHAASO experiment also detected \textasciitilde0.23° extended emission from the source, 1LHAASO J0007+7303u, and measured its spectrum up to ultra-high energies \cite{2013ApJ...764...38A}.

However, there are significant discrepancies between the spectral results published by VERITAS and LHAASO. These differences could be caused by a complex, energy-dependent source morphology that is not fully accounted for in the analyses. Understanding these discrepancies is essential for building an accurate physical model of the source.

To help clarify the VHE emission from CTA~1, we performed dedicated observations with the SST-1M telescopes. The SST-1M is a pair of small-sized imaging atmospheric cherenkov telescopes located at the Ondřejov Observatory in Czech Republic, capable of stereo observations \cite{2025JCAP...02..047A}. The system is sensitive to gamma rays in the energy range from 1 to 300 TeV. This work presents the preliminary results from 55 hours of stereo data, which aim to further constrain the source's characteristics and help resolve the tension between previous measurements.

\section{Observations and data selection}

The data presented in this work were collected between September 2024 and February 2025, resulting in a total of approximately 55 hours of observations for both telescopes.
The \texttt{sst1mpipe} package \cite{sst1mpipe_073} is a pipeline developed for the data reduction, calibration and analysis for the SST-1M telescopes. As described in ~\cite{2025arXiv250601733A}, it uses the standard random forest (RF) method for gamma-ray astronomy in order to perform the reconstruction of the events. The images from each telescope are calibrated and cleaned, before being parametrized according to the Hillas parametrization \cite{1985ICRC....3..445H}. RF then reconstruct the energy, direction, and type of the primary particle that initiated the air shower. This step is fundamental to gamma-ray astronomy, as it enables the extremely high hadronic background to be filtered out. The high-level analysis is then performed using the \texttt{gammapy} package \cite{gammapy:2023}.

As CTA~1 is a relatively faint source ($F_{>8 TeV}=4.9*10^{-12}$ erg cm$^{-2}$ s$^{-1}$, \cite{2013ApJ...764...38A}), a gamma efficiency of 50\% was chosen for this analysis, to exclude more background, with cuts optimized for the best signal to noise ratio for hard and faint sources. Moreover, to ensure accurate event reconstruction, it is crucial to have good data Monte-Carlo (MC) agreement. Following the works of \cite{2025arXiv250601733A, ICRC25Crab}, the event rate as a function of intensity was compared to the one of MC using power-law fit. In the selected sample, the event rate ratio to the MC is required to be in the range of 0.8 to 1.2. Finally, observations were selected to have below 700 MHz of night sky background (NSB) rate per pixel. After data quality cuts, the effective observation time for the stereo system is of 30h, partly due to unfavorable weather conditions. 
% With this dataset, a significance of 3.5$\sigma$ is reached in stereo (Figure \ref{fig:sig}), allowing the setting of upper limits to constrain the source's spectral properties.

\begin{figure}
    \centering
    \includegraphics[width=15cm]{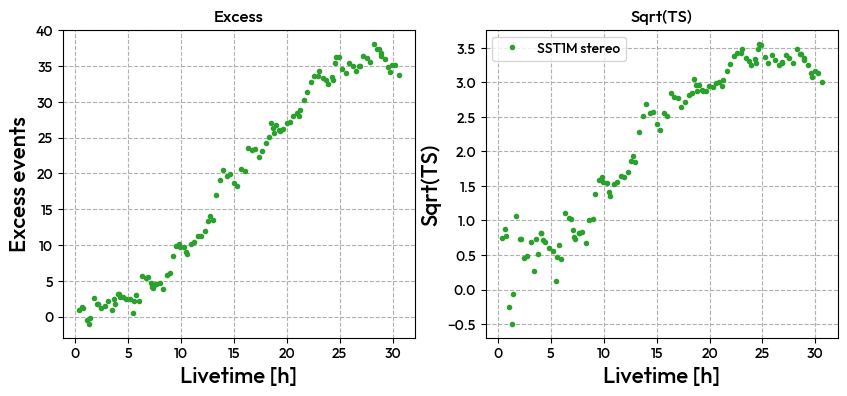}
    \caption{Excess and significance as a function of observation time for the selected stereo dataset, in a signal region of 0.3° radius.}
    \label{fig:sig}
\end{figure}

\section{Results}

\begin{figure}
    \centering
    \includegraphics[width=15cm]{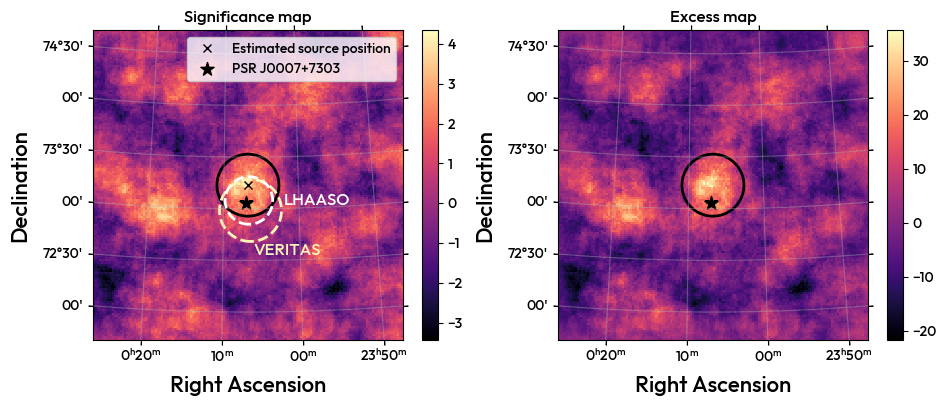}
    \caption{Sky maps of the CTA~1 region using the ring background estimation method. The black circle represents the region of the analysis. The black star is the position of PSR J0007+7303 \cite{2012ApJ...744..146A}. (\textbf{Left}) : Significance map. The white dashed circle is the LHAASO signal region and the beige one the VERITAS signal regions. (\textbf{Right}) : Excess map of the same region.}
    \label{fig:skymap}
\end{figure}
\begin{figure}
    \centering
    \includegraphics[width=7cm]{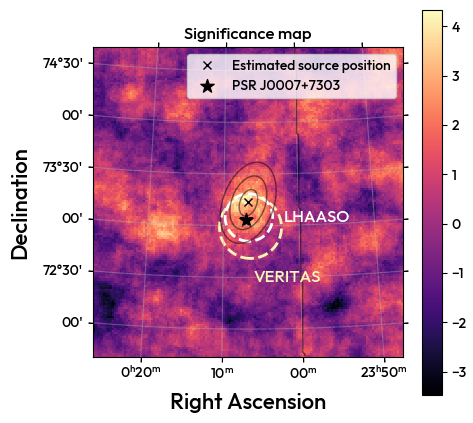}
    \caption{Significance map of the CTA~1 region using the FoV background estimation method. The contours represent the 2D gaussian fit performed to lecate the excess. The black star is the position of PSR J0007+7303.\cite{2012ApJ...744..146A}. The white dashed circle is the LHAASO signal region and the beige one the VERITAS signal regions.}
    \label{fig:skymap_fov}
\end{figure}

The analysis of 30 hours of stereo SST-1M data provides 3.5$\sigma$ evidence of gamma-ray emission from a region spatially coincident with the composite SNR CTA~1 (Figure~\ref{fig:sig}). This is a pre-trial significance with a power law (PL) spectrum with spectral index $\Gamma=2.0$. This observation aligns with the findings from both VERITAS and LHAASO, which firmly associate their detected gamma-ray signals, VER J0006+729 and 1LHAASO J0007+7303u respectively, with the PWN. 

For the morphological analysis shown in Figure \ref{fig:skymap}, the background is estimated using the ring background model described in \cite{2007A&A...466.1219B}. This technique uses a ring region centered on the test position, with radius 0.5° and width 0.3° in our case. This method is robust against gradients in the field of view (FoV), providing a reliable estimation of the hadronic background level. Moreover, it provides a background estimate for any point in the FoV, useful for locating a source like in the case of this work. Unlike the reflected background regions, the acceptance cannot be assumed constant, and an acceptance correction must be applied along the ring. For cross-checking purposes, the FoV background estimation method is also implemented to see if the excess is just a fluctuation. The background is estimated over the whole field of view outside of the exclusion mask. Finally, the reflected background regions method is used to compute the excess in the region of interest, where regions of equal acceptance are arranged around the pointing position to compute the background counts.

However, our result hint a new feature: the \textasciitilde3.5$\sigma$ excess is spatially offset by approximately 0.25° north of the pulsar (RA = 1.69°, Dec = 73.23°, displayed with the black cross in Figure \ref{fig:skymap}). 
This claim is strengthened by a similar offset using the FoV background estimation method on Figure \ref{fig:skymap_fov}.
This contrasts with the VERITAS and LHAASO analyses, which centered their morphological fits very near to the pulsar's position. This spatial discrepancy is the key to interpreting of our spectral results.

\begin{figure}
    \centering
    \includegraphics[width=8cm]{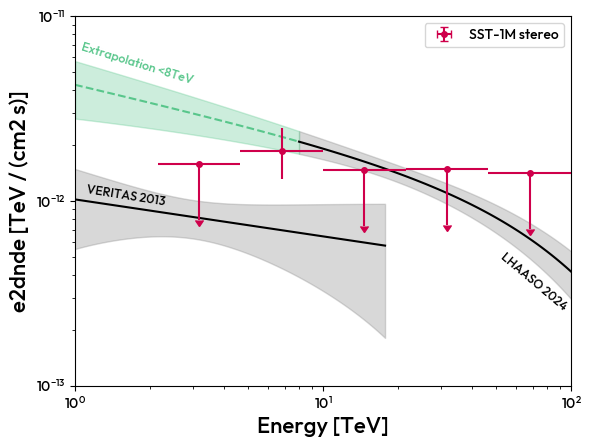}
    \caption{The differential spectrum of CTA 1. The red data point and upper limits are from this work (SST-1M stereo). The grey band represents the power-law model from VERITAS (2013), while the black line and green band show the power-law with exponential cutoff model from LHAASO (2024) and its extrapolation.}
    \label{fig:spec}
\end{figure}

Figure \ref{fig:spec} shows the SST-1M differential energy spectrum derived from this northern offset region, compared with the published results from VERITAS and LHAASO. It was produced using \texttt{gammapy} and the \texttt{FluxPointsEstimator}. It refits the normalization of the PL with fixed spectral index in each energy bin according to the bin-by-bin liklyhood method, notably used in \cite{2015ApJS..218...23A, 2018A&A...612A...1H}, and is suited for the low-statistics regime in which this work resides.

The VERITAS spectrum, measured between \textasciitilde0.6 TeV and 17.8 TeV, is described by a single power law with a spectral index of $\Gamma=2.2\pm0.2$. Our single flux point with $\sqrt{TS}>2$ at \textasciitilde8 TeV lies substantially above the VERITAS power-law model. Given that our emission originates from a different region, this suggests that the TeV flux is not uniform across the nebula.

The LHAASO experiment measured the spectrum from 8 TeV up to ~300 TeV. Their data cannot be described by a simple power law and are best fit by a power-law with an exponential cutoff (PLC), with an index $\Gamma=2.31\pm0.11$ and a cutoff energy $E_c=110\pm25$~TeV. Our flux point at \textasciitilde8 TeV is fully compatible with the LHAASO best-fit model. Furthermore, the upper limits we derive at energies above 20 TeV are consistent with VERITAS and with the spectral turnover reported by LHAASO and begin to constrain their model at the highest energies.

The fact that our spectrum, derived from a northern region, differs from the VERITAS spectrum (slightly southern region) but aligns with the higher-energy LHAASO spectrum motivates further investigation. It would imply that the region of the nebula we have detected may be dominated by higher-energy particles. LHAASO's own data supports this picture of an energy-dependent size, with the emission radius shrinking from 0.23° in the 8-100 TeV band to 0.17° for energies above 100 TeV. This behavior is characteristic of leptonic nebulae, where high-energy electrons cool faster and remain closer to their acceleration sites.

\section{Conclusion}

While these preliminary results are intriguing, the current statistical significance is insufficient to claim a definitive detection or to draw firm conclusions about the physical properties of the CTA~1 PWN. The hints of an offset emission region with a distinct spectrum are a strong motivation for continued study, but a deeper physical interpretation regarding particle transport mechanisms or magnetic field structures would be speculative at this stage.

To confirm the nature of this VHE emission, a deeper observation campaign with the SST-1M is essential and planned for the Summer 2025. The primary goals for the next observation cycle will be:

Increase statistical significance: Accumulate sufficient data to push the significance of the detection above the 5$\sigma$ threshold, which would provide unambiguous confirmation of the source.
Improve socalization and sorphology: With more data, and thanks to the \textasciitilde0.1° stereo angular resolution at 8-200~TeV of SST-1M \cite{2025arXiv250601733A}, we can better constrain the precise location and extension of the emission, confirming if the offset is a significant physical feature.
Measure a setailed spectrum: A stronger detection will allow us to measure the spectrum over a broader energy range.
Only with such improved statistics will it be possible to rigorously test the hypothesis of a complex, energy-dependent morphology. A confirmed, well-measured spectrum from this offset region would provide a crucial new input for physical models, helping to place meaningful constraints on the PWN's properties and ultimately resolving the puzzle of the discrepant results on CTA~1.

\section{Acknowledgements}
This publication was created as part of the projects funded in Poland by the Minister of Science based on agreements number 2024/WK/03 and DIR/\-WK/2017/12. The construction, calibration, software control and support for operation of the SST-1M cameras is supported by SNF (grants CRSII2\_141877, 20FL21\_154221, CRSII2\_160830, \_166913, 200021-231799), by the Boninchi Foundation and by the Université de Genève, Faculté de Sciences, Département de Physique Nucléaire et Corpusculaire. The Czech partner institutions acknowledge support of the infrastructure and research projects by Ministry of Education, Youth and Sports of the Czech Republic (MEYS) and the European Union funds (EU), MEYS LM2023047, EU/MEYS CZ.02.01.01/00/22\_008/0004632, CZ.02.01.01/00/22\_010/0008598, Co-funded by the European Union (Physics for Future – Grant Agreement No. 101081515), and Czech Science Foundation, GACR 23-05827S.

\bibliographystyle{JHEP}
\bibliography{bibliography.bib}

% \section{Full Autors List: SST-1M Collaboration}
% % \begingroup
% % \input{author_list.tex}
% % \endgroup

% \begingroup % Keep all changes local
% \setlength{\parindent}{0pt} % No indentation for this section

% % --- Author Block ---
% \def\authorsep{}
% % Redefine \author to include the superscript affiliation marker (#1)
% \renewcommand{\author}[2][]{\authorsep#2\textsuperscript{#1}\def\authorsep{, }\ignorespaces}

% % --- Affiliation Block ---
% \bigskip % Space between authors and affiliations
% % Redefine \affiliation to include its marker (#1) before the text
% \renewcommand{\affiliation}[2][]{\textsuperscript{#1} \textit{\small #2}\par}

% % --- Input your file ---
% % This file's content is now processed by the new definitions above
% \input{author_list.tex}

% \endgroup

%% Full authors list (ONLY FOR COLLABORATIONS)
\clearpage
\section*{Full Authors List: SST-1M Collaboration}
\scriptsize
\noindent
C.~Alispach$^1$,
A.~Araudo$^2$,
M.~Balbo$^1$,
V.~Beshley$^3$,
J.~Bla\v{z}ek$^2$,
J.~Borkowski$^4$,
S.~Boula$^5$,
T.~Bulik$^6$,
F.~Cadoux$^`$,
S.~Casanova$^5$,
A.~Christov$^2$,
J.~Chudoba$^2$,
L.~Chytka$^7$,
P.~\v{C}echvala$^2$,
P.~D\v{e}dic$^2$,
D.~della Volpe$^1$,
Y.~Favre$^1$,
M.~Garczarczyk$^8$,
L.~Gibaud$^9$,
T.~Gieras$^5$,
E.~G{\l}owacki$^9$,
P.~Hamal$^7$,
M.~Heller$^1$,
M.~Hrabovsk\'y$^7$,
P.~Jane\v{c}ek$^2$,
M.~Jel\'inek$^{10}$,
V.~J\'ilek$^7$,
J.~Jury\v{s}ek$^2$,
V.~Karas$^{11}$,
B.~Lacave$^1$,
E.~Lyard$^{12}$,
E.~Mach$^5$,
D.~Mand\'at$^2$,
W.~Marek$^5$,
S.~Michal$^7$,
J.~Micha{\l}owski$^5$,
M.~Miro\'n$^9$,
R.~Moderski$^4$,
T.~Montaruli$^1$,
A.~Muraczewski$^4$,
S.~R.~Muthyala$^2$,
A.~L.~Müller$^2$,
A.~Nagai$^1$,
K.~Nalewajski$^5$,
D.~Neise$^{13}$,
J.~Niemiec$^5$,
M.~Niko{\l}ajuk$^9$,
V.~Novotn\'y$^{2,14}$,
M.~Ostrowski$^{15}$,
M.~Palatka$^2$,
M.~Pech$^2$,
M.~Prouza$^2$,
P.~Schovanek$^2$,
V.~Sliusar$^{12}$,
{\L}.~Stawarz$^{15}$,
R.~Sternberger$^8$,
M.~Stodulska$^1$,
J.~\'{S}wierblewski$^5$,
P.~\'{S}wierk$^5$,
J.~\v{S}trobl$^{10}$,
T.~Tavernier$^2$,
P.~Tr\'avn\'i\v{c}ek$^2$,
I.~Troyano Pujadas$^1$,
J.~V\'icha$^2$,
R.~Walter$^{12}$,
K.~Zi{\c e}tara$^{15}$ \\

\noindent
$^1$D\'epartement de Physique Nucl\'eaire, Facult\'e de Sciences, Universit\'e de Gen\`eve, 24 Quai Ernest Ansermet, CH-1205 Gen\`eve, Switzerland.
$^2$FZU - Institute of Physics of the Czech Academy of Sciences, Na Slovance 1999/2, Prague 8, Czech Republic.
$^3$Pidstryhach Institute for Applied Problems of Mechanics and Mathematics, National Academy of Sciences of Ukraine, 3-b Naukova St., 79060, Lviv, Ukraine.
$^4$Nicolaus Copernicus Astronomical Center, Polish Academy of Sciences, ul. Bartycka 18, 00-716 Warsaw, Poland.
$^5$Institute of Nuclear Physics, Polish Academy of Sciences, PL-31342 Krakow, Poland.
$^6$Astronomical Observatory, University of Warsaw, Al. Ujazdowskie 4, 00-478 Warsaw, Poland.
$^7$Palack\'y University Olomouc, Faculty of Science, 17. listopadu 50, Olomouc, Czech Republic.
$^8$Deutsches Elektronen-Synchrotron (DESY) Platanenallee 6, D-15738 Zeuthen, Germany.
$^9$Faculty of Physics, University of Bia{\l}ystok, ul. K. Cio{\l}kowskiego 1L, 15-245 Bia{\l}ystok, Poland.
$^{10}$Astronomical Institute of the Czech Academy of Sciences, Fri\v{c}ova~298, CZ-25165 Ond\v{r}ejov, Czech Republic.
$^{11}$Astronomical Institute of the Czech Academy of Sciences, Bo\v{c}n\'i~II 1401, CZ-14100 Prague, Czech Republic.
$^{12}$D\'epartement d'Astronomie, Facult\'e de Science, Universit\'e de Gen\`eve, Chemin d'Ecogia 16, CH-1290 Versoix, Switzerland.
$^{13}$ETH Zurich, Institute for Particle Physics and Astrophysics, Otto-Stern-Weg 5, 8093 Zurich, Switzerland.
$^{14}$Institute of Particle and Nuclear Physics, Faculty of Mathematics and Physics, Charles University, V Hole\v sovi\v ck\' ach 2, Prague 8, Czech~Republic.
$^{15}$Astronomical Observatory, Jagiellonian University, ul. Orla 171, 30-244 Krakow, Poland.

\end{document}